# Fourier holographic endoscopy for label-free imaging through a narrow and curved passage

Wonjun Choi[1,2,+], Munkyu Kang[1,2,+], Jin Hee Hong[1,2], Ori Katz[3], Youngwoon Choi[4,*], and Wonshik Choi[1,2,*]

[1]Center for Molecular Spectroscopy and Dynamics, Institute for Basic Science (IBS), Seoul 02481, Republic of Korea
[2]Department of Physics, Korea University, Seoul 02481, Republic of Korea
[3]Department of Applied Physics, The Selim and Rachel Benin School of Computer Science & Engineering, The Hebrew University of Jerusalem, Jerusalem 9190401, Israel
[4]Department of Bioengineering, Korea University, Seoul 02481, Republic of Korea
+These authors contributed equally to this work.
*e-mail: youngwoon@korea.ac.kr and wonshik@korea.ac.kr

## Abstract

Ultrathin lensless fibre endoscopes offer minimally invasive investigation, but they mostly operate as a rigid type due to the need for prior calibration of a fibre probe. Furthermore, most implementations work in fluorescence mode rather than label-free imaging mode, making them unsuitable for medicine and industry. Herein, we report a fully flexible ultrathin fibre endoscope taking 3D holographic images of unstained tissues with 0.87-μm spatial resolution. Using a bare fibre bundle as thin as 200-μm diameter, we design a lensless Fourier holographic imaging configuration to selectively detect weak reflections from biological tissues, a critical step for stain-free reflectance imaging. A unique algorithm is developed for calibration-free holographic image reconstruction, allowing us to image through a narrow and curved passage regardless of fibre bending. We demonstrate endoscopic reflectance imaging of unstained rat intestine tissues that are completely invisible to conventional endoscopes. The proposed endoscope will expedite more accurate and earlier diagnosis than before with minimal complications.

Optical microscopy is an essential tool for understanding the physiology of living tissues due to its high spatial resolution, molecular specificity, and minimal invasiveness[1]. However, these benefits are out of reach when target objects are located either inside curved passages or underneath light-scattering tissues. By visualizing these hard-to-reach areas, endoscopes have revolutionized medical practice for early disease diagnosis. Over the past decade, endoscopes with microscopic resolution have been developed for more accurate and earlier diagnosis[2,3]. In addition, the demand for ultrathin endoscopes (with a probe diameter well below 1 mm) has been steadily growing to minimize the discomfort and complications accompanied by insertion of the endoscope probe[3–7].

Endoscopic microscopy typically employs various optical fibres as thin and flexible light-guiding channels. For example, a single optical fibre has been used by attaching various types of scanning devices and optical elements to the distal side of the fibre facing the sample[4,9–11]. Multiphoton imaging[4,10,12] and optical coherent tomography (OCT)[8,13,14] have been implemented in this configuration. However, the scanner attached to the fibre is often too bulky to be ultrathin even though the diameter of the fibre itself is small. Image guiding media such as coherent fibre bundles are used to eliminate the need for distal scanners, thus making the endoscope probe thinner and more robust. Individual fibre cores in the bundle are used as image pixels by either attaching imaging optics to the tip of the fibre bundle or making direct contact of the fibre tip to the sample surface[15,16]. Wide-field fluorescence imaging modality has been implemented for rapid medical diagnosis[17,18]. And confocal fluorescence imaging has been realized by the high-speed scanning of the focus at the proximal side of the fibre outside the subject[15,19]. One critical drawback of this configuration is its inability to acquire label-free reflectance images of biological tissues. Strong back-reflection of the illumination occurring at the fibre cores exactly coincides with much weaker reflection signals from the biological tissues. This is one of the main reasons why fluorescence imaging mode is widely used, in which fluorescence emission can easily be separated from the back-reflection noise by using colour filters. A simple solution for intrinsic reflectance imaging is to introduce a separate fibre for the illumination. However, this has only been applicable for macroscopic imaging due to the enlarged probe diameter and low spatial resolving power constrained by the fibre-sample distance required for the separate illumination.

Various types of lensless fibre endoscopes have been developed to minimize the probe size to the diameter of fibre itself. For example, multimode optical fibres have been employed as image guiding medium[20–22]. They provide numerous independent spatial modes that can transfer image information at once. Since their areal mode density is one or two orders of magnitude larger than the fibre core density in the fibre bundle, the probe diameter can be even smaller than the fibre bundle probe. However, complex mode mixing obscures image information. Many recent studies have proved that the prior calibration of a multimode fibre in the form of a transmission matrix (TM) enables the recovery of the object image[20,21,23,24]. This concept has been applied to fluorescence endoscopic imaging of mouse brain[25,26] and wide-field reflectance imaging of biological tissues[22]. The TM approach has also been applied to a fibre bundle endoscope for either fluorescence[24] or transmittance/reflectance imaging[27–29] for the removal of both pixelation artefacts and distal optics. However, TM-based endoscopes have so far been used as a rigid type rather than a flexible type[25,26], mainly because the TM calibration becomes invalid when the fibres are bent or twisted during insertion. Several approaches have been introduced to resolve this critical constraint. For example, speckle correlation due to the memory effect of the fibre bundle has been used to reconstruct object image without prior knowledge of the TM[30,31]. Feedback optimization of two-photon signals measured at the proximal end allows direct access to the fibre TM[32], and a virtual guide star is administrated to the distal side to dynamically compensate for fibre bending[33,34]. The use of specially designed fibres that have less sensitive to the bending deformation has been proposed[35–37]. While these studies have shown the feasibility to a certain extent, innate limitations such as the lengthy optimization time, enlargement of the probe unit, constraint for the bending configurations, and difficulty in fabricating ideal fibres still preclude the realization of the flexible endoscopes. Essentially, previous lensless fibre endoscopes either require prior calibration or operate in fluorescence imaging mode, and an ideal endoscope satisfying all the critical requirements – flexibility, ultrathin probe diameter, microscopic resolution and stain-free imaging – has yet to be developed.

Herein, we present a fully flexible and ultrathin fibre endoscope that can perform high-resolution holographic imaging of unstained tissues through an extremely narrow and curved passage. Our endoscope probe is simply a bare fibre bundle with no lens or scanner attached to the tip. Therefore, the diameter of the endoscope is equal to that of the fibre bundle itself, which was either 200 μm or 350 μm in the present study. Unlike a conventional fibre bundle endoscope, the fibre cores are not image pixels. Instead, the fibre is placed at any distance larger than 400 μm from the target sample. In this unique configuration, the illumination is made through each fibre core, one by one, but the detection is made by all the other fibre cores. The reflected waves from the sample that are blurred and spread to the other fibre cores are measured by using a holographic field measurement[29]. This allows us to selectively block the back-reflection noise from the illumination core. Since the measured holographic field contains the Fourier transform of an object function by means of Fresnel diffraction, we term our method 'Fourier holographic endoscopy'. However, an object image cannot be reconstructed from the detected field due to the dynamically varying core-to-core phase retardations with respect to the fibre bending and twisting. While the fibre was calibrated before its insertion in previous TM approaches, we developed an algorithm that can identify the complex core-to-core phase retardations and reconstruct the diffraction-limited and pixelation-free object image all at the same time without the need for any prior calibration. This allowed us to perform endoscopic imaging with an arbitrary and varying bending configuration. For an object located behind a narrow and curved passage, we demonstrated the endoscopic reflectance imaging with a microscopic spatial resolution of 0.87 μm for low-contrast resolution targets and rat intestine tissues that are completely invisible to conventional endoscopes. Furthermore, we realize the volumetric 3D imaging covering the depth range of 400–1200 μm with an axial resolution as good as 7.5 μm by exploiting the holographic image reconstruction.

**Experimental layout of Fourier holographic endoscopy**

Our experimental setup is shown in Fig. 1a. A diode laser (Finesse Pure, Laser Quantum) with a wavelength of $\lambda = 532$ nm and a coherence length of 6 mm was used as the light source. The output beam was split into sample and reference waves with a beam splitter (BS1). The sample beam was reflected by a two-axis galvanometer mirror (GM) and focused on the input plane (IP) of a coherent fibre bundle via an objective lens (OL). Two types of fibre bundles were used, one with 350 μm diameter (Fujikura, FIGH-10-350S) and the other with 200 μm diameter (Fujikura, FIGH-10-200S). The 350-μm-diameter fibre bundle (1 m-long, and a minimum bending radius of 35 mm) contains around 10,000 fibre cores, the diameter of each being around 2 μm. We adjusted the GM scanning angle to sequentially focus the illumination beam on each fibre core. This ensured that the illumination beam (indicated as green) travelled through the single fibre core and came out from the output plane (OP) of the fibre bundle via the same fibre core.

To separate the back-reflection noise from the fibre bundle, we placed the sample plane (SP) at a standoff distance $d$ from the OP. This allowed the backscattering signals from the sample to be spread in space and collected at fibre cores other than the illumination core. To be demonstrated later, such a standoff distance does not sacrifice imaging resolution. In fact, it can provide better spatial resolution than what can be achieved in the conventional manner when the fibre is in contact with the sample. Typically, $d$ ranged from 400 to 1200 μm to ensure that the Fresnel approximation is valid[38]. In this arrangement, the illumination beam at the OP arrives at the SP as a parabolic wave. The angle of the parabolic illumination wave at the SP is varied depending on the position of the illumination core at the IP. The waves backscattered by the object at SP (indicated as yellow) are captured by multiple fibre cores at the OP and delivered back to the IP. The waves returning through the IP are captured by the OL and delivered to the camera. To obtain the amplitude and phase of the returning waves, a reference wave was introduced at the camera with an off-axis configuration, and the Hilbert transform was applied to the recorded holograms.

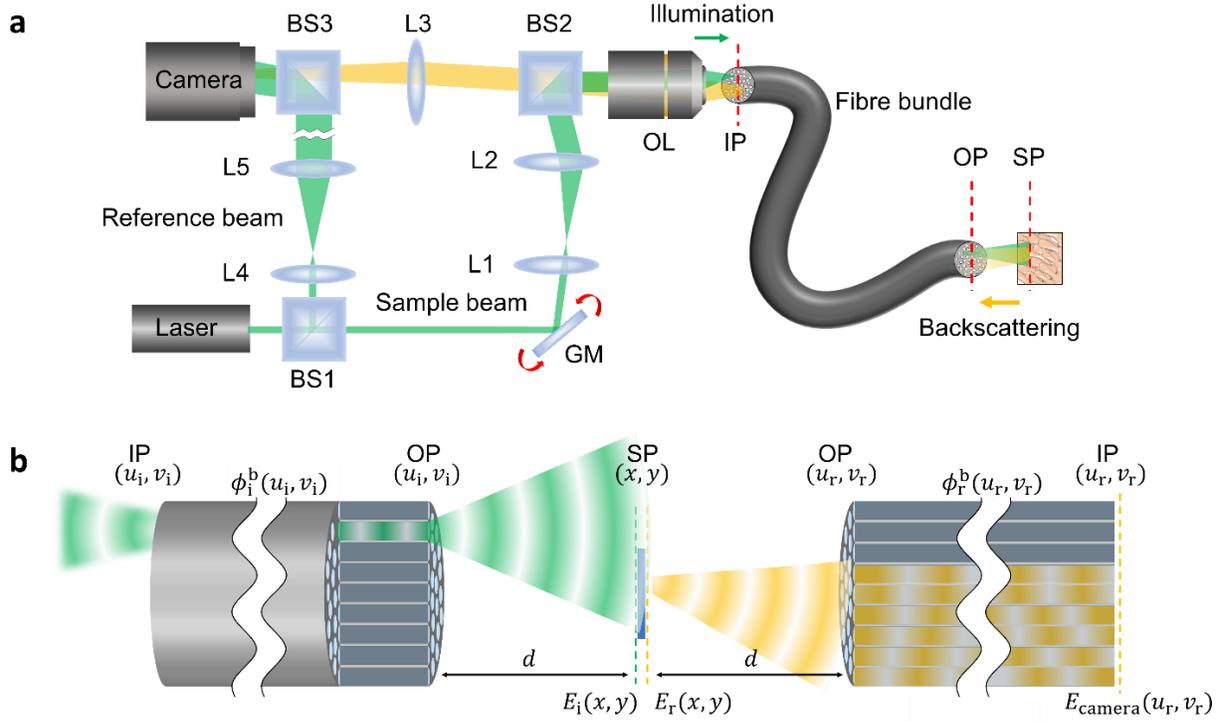

**Fig. 1. Schematic of the experimental setup and the image formation principle. a,** Experimental setup. The output beam from a laser is divided into sample and reference beams. The sample beam is delivered to the sample through the fibre bundle. The backscattering signal from the sample, indicated as yellow for clarity although its wavelength is identical to the incident wave, is captured by the fibre bundle and delivered to the camera. The reference beam generates an interferogram together with the signal beam at the camera. GM: 2-axis galvanometer scanning mirror, BS1-3: beam splitters, L1-5: lenses, OL: objective lens. IP and OP: input and output planes of the fibre bundle, respectively. SP: sample plane. **b,** Image formation principle. The illumination and reflection pathways have been unfolded to make their distinction clear. $(u_i, v_i)$ and $(u_r, v_r)$: spatial coordinates at the fibre bundle for illumination and detection pathways, respectively; $\phi_i^b(u_i, v_i)$ and $\phi_r^b(u_r, v_r)$: phase retardations induced by the fibre bundle during the illumination and reflection, respectively; $E_i(x,y)$ and $E_r(x,y)$: the electric fields incident to and reflected by the sample, respectively; and $E_{camera}(u_r, v_r)$: the electric field detected at the camera..

## Calibration-free image reconstruction principle

The detailed layout of the image formation process is shown in Fig. 1b. The incident wave focused at the fibre core located at $(u_i, v_i)$ of the IP experiences phase retardation $\phi_i^b(u_i, v_i)$ at the corresponding core and exits as a point source at the OP. This point source travels to the SP and meets the target object as a parabolic wave. The incident wave at $(x,y)$ of the SP can be written as

$$E_i(x,y; u_i, v_i) = \frac{e^{ikd}}{i\lambda d} exp\left\{i\frac{k}{2d}[(x-u_i)^2 + (y-v_i)^2]\right\} e^{i\phi_i^b(u_i,v_i)}. \quad (1)$$

Here, $k = 2\pi\lambda^{-1}$ is the wavenumber. The wave is reflected by the target sample having object function $O(x,y)$, which is the amplitude reflectance of the target. The reflected wave given by $E_r(x,y; u_i, v_i) = O(x,y)E_i(x,y; u_i, v_i)$ is depicted on the transmission side of the sample in Fig. 1b to make it clearly distinct from the incident wave. The reflected wave travels back to $(u_r, v_r)$ of the OP following Fresnel diffraction and returns to the IP after experiencing fibre core-dependent phase retardation $\phi_r^b(u_r, v_r)$. After incorporating all of these processes, we can obtain the field at the camera, i.e. the field detected at the fibre proximal facet plane, as

$$E_{\text{camera}}(u_r, v_r; u_i, v_i) = -\frac{e^{2ikd}}{\lambda^2 d^2} e^{i\phi_r(u_r, v_r)} \tilde{O}_M\left(\frac{k}{d}(u_r + u_i), \frac{k}{d}(v_r + v_i)\right) e^{i\phi_i(u_i, v_i)}. \quad (2)$$

Here, $\tilde{O}_M$ is the Fourier transform of the modified object function $O_M(x, y) = O(x, y) exp\left\{i\frac{k}{d}(x^2 + y^2)\right\}$, and $\phi_i(u_i, v_i) = \frac{k}{2d}(u_i^2 + v_i^2) + \phi_i^b(u_i, v_i)$ and $\phi_r(u_r, v_r) = \frac{k}{2d}(u_r^2 + v_r^2) + \phi_r^b(u_r, v_r)$ are the phase retardations in the illumination and reflection pathways, respectively. In essence, the measured field contains the object spectrum with scaling factor $k/d$ and spectral shift by $(-u_i, -v_i)$. The quadratic phase terms are introduced because of Fresnel diffraction. Interestingly, they are separately absorbed in $O_M(x, y)$, $\phi_i(u_i, v_i)$, and $\phi_r(u_r, v_r)$.

Finding $O_M(x, y)$, $\phi_i(u_i, v_i)$, and $\phi_r(u_r, v_r)$ from the measured reflection waves is generally a difficult task, especially when the phase retardations among the cores are uncorrelated. However, Eq. (2) is formally equivalent to the imaging configuration where the target object embedded within an aberrating medium is located at the exact focal plane of the objective lens [39]. In our previous studies [39–41], we developed a method termed 'closed-loop accumulation of single scattering (CLASS)' to separate the object function from complex sample-induced aberrations and recover the ideal diffraction-limited spatial resolution. We specially redesigned this CLASS algorithm to address the present problem by accounting for the pixelated sampling by the fibre bundle and the quadratic phase terms. The working principle is to first construct the reflection matrix $\boldsymbol{R}$ in which the elements are filled with $E_{\text{camera}}(u_r, v_r; u_i, v_i)$ with $(u_i, v_i)$ and $(u_r, v_r)$ as column and row indices, respectively. We then iteratively compute the correlations among the columns and rows to identify $\phi_i(u_i, v_i)$ and $\phi_r(u_r, v_r)$, from which we obtain $O_M(x, y)$ (see Methods for details of image reconstruction). We can obtain the reflectance map of the target sample by the simple relationship $|O_M(x, y)|^2 = |O(x, y)|^2$. This algorithm automatically identifies $\frac{k}{d}(x^2 + y^2)$ in $O_M(x, y)$ for the objects within working range. Since the fibre-to-sample distance $d$ is automatically obtained from the quadratic phase terms, it is also possible to reconstruct a 3D image from the recording of a single reflection matrix.

**Procedures for data acquisition and image reconstruction**

Experimental procedures to perform the endoscopic imaging are shown in Fig. 2. We placed a US Air Force (USAF) resolution target (Edmund, 2" x 2" Positive, 1951 USAF Hi-Resolution Target #58-198) at the SP. The conventional full-field endoscopic image taken by the same fibre bundle is shown in Fig. 2h as a point of reference (see Methods for details). From the photograph of the fibre bundle surface taken at the IP, we chose which fibre cores to illuminate (the red dots in Fig. 2a). In an arbitrary fibre bending configuration, we adjusted the GM angle to focus the illumination beam on each core at a time. In a typical experiment, illumination was made to 100–3000 different fibre cores. Raw images of the backscattered waves from the sample taken at the IP are shown in Fig. 2b for the illumination of a few representative core fibres. The spatial coordinates of these output images correspond to $(u_r, v_r)$ at the fibre bundle surface. In each image, the bright dot is due to the direct back-reflection from the fibre core where the illumination was focused. They were moving synchronously with the scanning of $(u_i, v_i)$. Backscattered signal waves by the target object were spatially spread over the other cores in the recorded images. We selectively removed the back-reflection noise by dropping the signal at the pixel where the back-reflection noise was focused. We then performed the Hilbert transform on the raw images in Fig. 2b to obtain a complex field map of the backscattered waves (Fig. 2c).

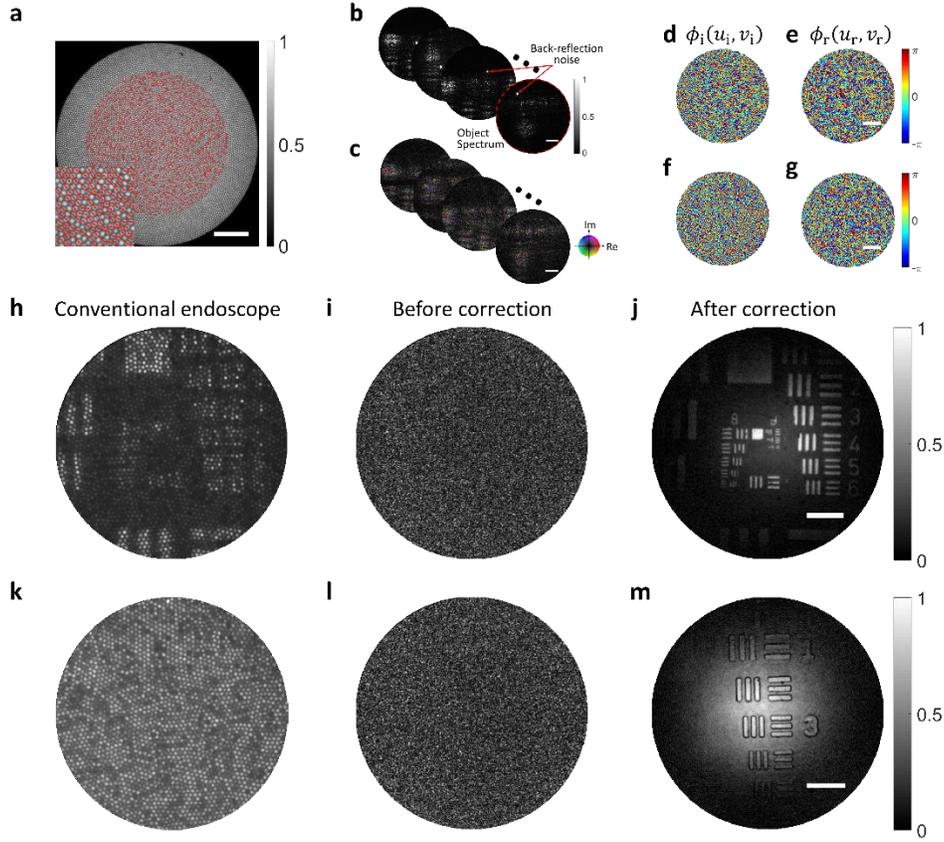

**Fig. 2. Data acquisition and image reconstruction. a** Bright-field image of the fibre bundle taken at the IP by illuminating the incoherent source from the OP. The fibre cores where the illumination beam was focused are indicated as red dots. **b,** Raw images captured by the camera for the scanning of $(u_i, v_i)$. For better visibility, the images shown here were taken without the reference beam, while the interference images were recorded for the endoscopic imaging. The colour bar indicates normalized intensity excluding the back-reflection noise. **c,** Complex field maps $E_{\text{camera}}(u_r, v_r; u_i, v_i)$ obtained from the raw interference images in **b**. Circular colour map: real and imaginary values of $E_{\text{camera}}$. **d** and **e,** Fibre core-dependent phase retardations $\phi_i(u_i, v_i)$ and $\phi_r(u_r, v_r)$, respectively, identified via the algorithm. The colour bar signifies the phase in radians. Scale bar, $0.1k$. **f** and **g,** The same as **d** and **e**, respectively, but for a low-contrast resolution target. **h,** Conventional endoscopic image of a USAF target taken by the incoherent illumination from the IP. The fibre bundle was in contact with the target surface. **i,** Coherent addition of inverse Fourier transformed images of the complex field maps in **c** before correcting for the fibre core-dependent phase retardations. **j,** The same as **i** but after the correction. The colour bar signifies normalized amplitude. **k-m,** The same as **h-j**, respectively, but for a low-contrast resolution target. The 350-μm-diameter fibre bundle was used, and the fibre-to-sample distance $d$ was 500 μm for **i**, **j**, **l**, and **m**. Scale bars in **a** to **c**: 50 μm. Scale bars in **j** and **m**: 30 μm.

The detected images in Fig. 2c correspond to $E_{\text{camera}}(u_r, v_r; u_i, v_i)$ in Eq. (2). Since $\phi_i^b(u_i, v_i)$ and $\phi_r^b(u_r, v_r)$ vary significantly during the insertion of the fibre, the object image cannot be reconstructed directly from the images in Fig. 2c. The inverse Fourier transform of the images in Fig. 2c resulted in speckled patterns, and their coherent summation showed no sign of image information (Fig. 2i). As described earlier, we constructed a reflection matrix from the images in Fig. 2c and applied our algorithm to identify $\phi_i(u_i, v_i)$ and $\phi_r(u_r, v_r)$, which are respectively shown in Fig. 2d and 2e. They presented extremely complex phase patterns, as was expected. We compensated for these core-dependent phase retardations to the images in Fig. 2c, which were then coherently added after accounting for the spectral shift by $(u_i, v_i)$ (see Methods for the algorithm and image reconstruction). The final reconstructed image shown in Fig. 2j presents fine details in Groups 8 and 9 that were invisible in the conventional endoscope image in Fig. 2h. The spatial resolution was as good as 0.87 μm since element #2 in Group 9 of the USAF was successfully resolved. Furthermore, there was no pixelation in the reconstructed image because the acquired images via $E_{\text{camera}}$ were pixelated in the spatial frequency but not in real space.

The image resolution of our endoscope is determined by the diameter of the fibre bundle $D$, which sets the numerical aperture (NA) by $\alpha = n(D/2)d^{-1}$, where $n$ is the refractive index of the medium between the fibre and the sample. When $\alpha$ is larger than the NA of the fibre itself (0.4), then the latter limits the achievable spatial resolution. As $d$ is reduced from 1.2 mm to 400 μm, $\alpha$ is increased from 0.12 to 0.47 for $n = 1$, and the theoretical spatial resolving power is increased from 1.6 to 0.67 μm. In the experiment, the reflected waves within the diameter of $D_{\text{eff}} = 0.80D$ were used for the sake of the computational time. Therefore, the diffraction-limited resolution at $d = 500$ μm was 0.86 μm, which is in good agreement with the experimental result. The view field diameter was set as $L = (\lambda/n)d\Delta D^{-1}$, where $\Delta D = 3.2$ μm is the fibre core-to-core spacing. Therefore, $L$ increased from 66 μm to 170 μm with an increase in $d$. In the image reconstruction process, the effective core-to-core spacing was reduced to $\Delta D_{\text{eff}} = 1.5$ μm due to the synthesis of multiple images with the spectral shift by $(u_i, v_i)$. As a result, $L$ ranged from 140 to 410 μm with an increase in $d$. The estimated view field at $d = 500$ μm was 170 μm, which agrees well with the experimental result.

To prove the benefit of our method in removing the back-reflection noise, we performed imaging of a low-contrast USAF resolution target (Thorlabs, Positive 1951 USAF test target, R1DS1P). The reflectance of the target was 10 % higher than that of the background at 532 nm wavelength. The conventional endoscopic imaging (Fig. 2k) did not show any structures at all because the back-reflection noise was much stronger than the backscattering signal from the resolution target. On the contrary, our method could clearly unveil the target structures (Fig. 2m). Fig. 2f and 2g show the identified core-dependent phase retardations.

**Endoscopic imaging through a narrow and curved passage**

We demonstrated label-free microscopic imaging through a narrow and curved passage. A target object was placed inside a box, and an 80-cm-long plastic tube was connected to a hole at the ceiling of the box (Figs. 3a and 3b). The inner and outer diameters of the tube were 4.3 and 6.5 mm, respectively. While the proximal end of the fibre was fixed at the focal plane of the OL (the inset in Fig. 3b), the distal side of the fibre was inserted through the tube until it reached the target (Edmund, 2" x 2" Positive, 1951 USAF Hi-Resolution Target #58-198). A conventional endoscopic image was taken when the fibre bundle was in contact with the target (Fig. 3c). Although this image was pixelated and its resolving power was low, it allowed us to position the fibre to the region of interest. Once the target was identified, we pulled back the fibre bundle to set $d$ within the working range and recorded a reflection matrix by scanning the illumination fibre cores. The acquired image is shown in Fig. 3d, in which the reconstructed image is autofocused to the target object and fine details of Group 9 are clearly visible with a spatial resolution of 1.1 μm. The standoff distance was calculated as 800 μm from the quadratic phase of the reconstructed image. This result confirms that our endoscope can take an *in-situ* microscopic image of a target object through a narrow and curved passage with no prior calibration of the fibre bundle.

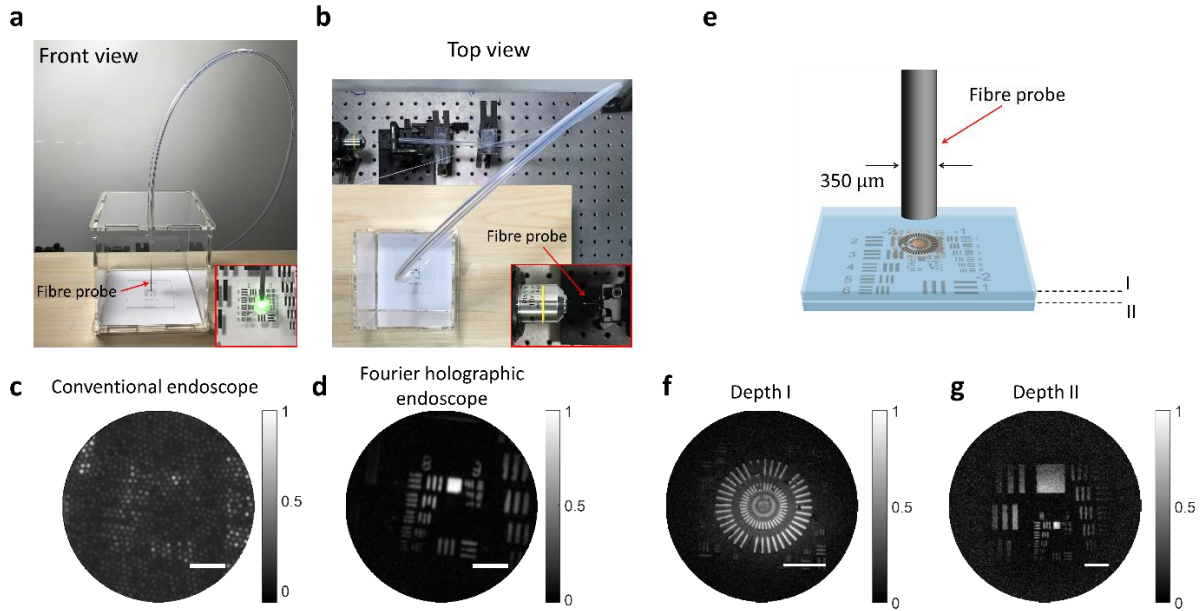

**Fig. 3. Endomicroscopic imaging through a narrow and curved passage, and 3D imaging capability. a** and **b**, Front and top views of the experimental configuration, respectively. The zoomed-in image in **a**: the distal side of the fibre bundle near the sample. The zoomed-in image in **b**: the light coupling to the IP of the fibre from the objective lens, respectively. **c** and **d**, Conventional endoscopic image and the reconstructed image with our endoscope, respectively. Scale bars: 20 μm. The colour bar in **c** shows normalized intensity and the colour bar in **d** shows normalized amplitude. **e,** Schematic for endoscopic imaging of stacked targets. Two resolution targets were placed at two different depths, I and II, with fibre-to-sample distances of 0.6 and 1.06 mm, respectively. **f** and **g**, Endoscopic images for the depths of I and II, respectively, reconstructed using a single reflection matrix recording. The 350-μm-diameter fibre bundle was used for image acquisition. Scale bars: 30 μm. The colour bar signifies normalized amplitude.

**3D image reconstruction from a single matrix recording**

Our endoscope has 3D imaging capability since our algorithm works as a holographic image reconstruction. For the two targets separated by 460 μm (Fig. 3e), we took a single reflection matrix. Our algorithm identified the target image from depth I (Fig. 3f) as the reflection images since this depth gave stronger correlations than depth II when computing the core-dependent phase retardations. Once $\phi_i(u_i, v_i)$ and $\phi_r(u_r, v_r)$ had been identified for depth I, we applied the computational propagation of each $E_{camera}(u_r, v_r; u_i, v_i)$ based on the angular spectrum method [38] after applying the correction to obtain the target image at depth II. After numerical refocusing, completely different structures corresponding to depth II came to light (Fig. 3g). The measured depth resolution of our endoscope image was as good as 7.5 μm.

**Microscopic imaging of unstained rat intestine tissues**

We demonstrated the endomicroscopic imaging of unstained biological tissues. An intestine tissue excised from a rat was placed in immersion oil (see Methods for the sample preparation). At first, the tip of the fibre bundle was placed close to the surface of the villi in the intestine. The image acquired by the conventional endoscope is shown in Fig. 4a, in which the contrast between the target and background was too low to discern any villus. In fact, the contrast of the villi was so low that they could be vaguely visible even in the transmission image (Fig. 4b) taken by illuminating from the sample side. We then pulled back the fibre bundle about 1000 μm away from the villi and took endoscopic images for different regions. Representative images obtained for the circular areas indicated in Fig. 4b are shown in Figs. 4c to 4f. The boundary of the villi is sharply resolved with a much better contrast than the conventional endoscope image, and the structures of the relatively transparent inner parts near the boundary are recognizable. The entire map of two villi was reconstructed by stitching multiple images taken for different regions (Fig. 4g). The spatial resolution obtained from the sharpness of the boundary was about 1 μm, far better than typical colonoscopy can provide. This result suggests that our endoscope's label-free microscopic imaging capability can potentially be used for identifying the abnormalities occurring at the surface of the villi at much earlier stage than the current practice.

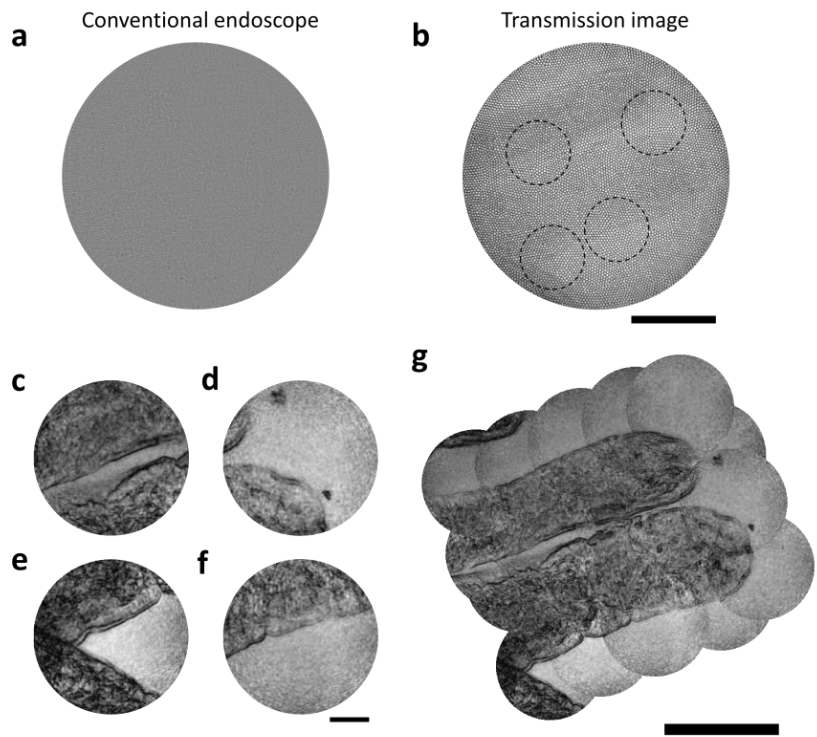

**Fig. 4. Microscopic imaging of villi in a rat intestine. a,** Conventional reflectance endoscope image taken when the fibre bundle was in contact with the villi. **b,** Conventional transmission image obtained through the fibre bundle. The LED illumination was sent from the villi to the fibre bundle. Scale bar: 100 μm. **c-f,** Label-free reflectance images obtained using our holographic endoscope for the circular areas indicated in **b**. Scale bar: 25 μm. **g,** Reconstructed image of two villi by stitching multiple images taken over wide region of interest. The 350-μm-diameter fibre bundle was used for image acquisition. Scale bar: 100 μm.

**Discussion**

We developed a fully flexible ultrathin lensless endoscope that can perform microscopic imaging of unstained biological tissues through a narrow and curved passage. The proposed method takes the thinnest possible form because the fibre bundle itself is used as the endoscope probe. The diameter of the probe was either 350 μm or 200 μm in the present study, but it can be reduced further just by employing a smaller fibre bundle. Therefore, the diameter can even be comparable to the thinnest acupuncture needle. The achieved lateral and axial resolutions were 0.87 and 7.5 μm, respectively, which are comparable to those of high-resolution microscopes. The recording of holographic images enabled the reconstruction of 3D objects for a volume covering a depth range of 400–1200 μm using a single reflection matrix recording. While retaining all of these benefits, our endoscope operates as a flexible type rather than a rigid type because we eliminated the need for prior calibration of the imaging probe. We resolved the complex core-dependent phase retardations directly from the images taken during the imaging session, which allowed us to freely navigate the endoscope probe around the region of interest. The acquisition time for the volumetric imaging was as short as 1s with the use of an sCMOS camera because the required complex field maps can be as small as 100 images, and it can be reduced to less than a second with the use of a high-speed camera.

The capability of label-free and flexible endoscopic imaging with a microscopy level of performance could expedite non-invasive or minimally invasive disease diagnosis and industrial inspections. For example, our endoscope is so thin that it can be inserted through extremely narrow cavities such as the minute airways in the lung and narrow microvessels for the microscopic investigation. Since the probe can be as thin as an acupuncture needle, it can directly be inserted into the brain tissues to find the neurological disorders with minimal complications. The applicability is not limited to the medical diagnosis but can be extended to the industrial inspections. In modern semiconductors and microprocessors, devices are stacked in layers for maximal integration. Our endoscope can be used to monitor each step of fabrication process occurring inside the chambers with minimal interventions. Label-free imaging capability is particularly well-suited because the use of chemical staining is not desirable due to the concern about contaminating the devices.

## Methods

**Correcting the fibre core-dependent phase retardations.** From complex images $E_{\text{camera}}(u_r, v_r; u_i, v_i)$ in Fig. 2c, we constructed reflection matrix $\boldsymbol{R}$ in which the column and row indices correspond to $(u_i, v_i)$ and $(u_r, v_r)$, respectively. This was achieved by converting the individual images in Fig. 2c into column vectors and appending them together to form a matrix. We identified $\phi_i^c(u_i, v_i)$ from the correlation among the columns and constructed a corrected matrix in which the matrix elements were $E^c(u_r, v_r; u_i, v_i) = e^{-i\phi_i^c(u_i,v_i)} E_{\text{camera}}(u_r, v_r; u_i, v_i)$. We then identified $\phi_r^c(u_r, v_r)$ from the correlation among the rows of the corrected matrix, which was then applied to construct the corrected matrix whose elements were $E^c(u_r, v_r; u_i, v_i) = e^{-i\phi_i^c(u_i,v_i)} E_{\text{camera}}(u_r, v_r; u_i, v_i) e^{-i\phi_r^c(u_r,v_r)}$. These processes were repeated until the object spectrum $\tilde{O}_M\left(\frac{k}{d}(u_r + u_i), \frac{k}{d}(v_r + v_i)\right)$ was identified[39]. We shifted the object spectrum by $(u_i, v_i)$ for individual images, which converted the acquired spectrum to $\tilde{O}_M\left(\frac{k}{d}u_r, \frac{k}{d}v_r\right)$ for all choices of illumination cores. These spectra were added together, and an inverse Fourier transform was taken to obtain coherently accumulated object function $O_M(x, y)$. $|O_M(x, y)| = |O(x, y)|$ is shown in Figs. 2j and 2m.

**Acquisition of the conventional endoscope images.** Conventional endoscope images shown in Figs. 2h, 2k, 3c, and 4a were taken by bringing the tip of the fibre bundle into contact with the resolution target. This is equivalent to attaching a lens having the magnification of 1x to the tip of a fibre. A light-emitting diode (Model: M530L3, Thorlabs, wavelength: 530 nm) was inserted in the sample beam path in Fig. 1a and its output beam illuminated the entire fibre bundle. Reflected images were recorded with the same camera used for our endoscope imaging.

**Preparation of the rat intestines.** We extracted intestinal tissues from 1- to 2-day-old Sprague-Dawley rats. The upper part of the small intestine was quickly excised and fixed for 2–4 hours at 4 °C in 4% paraformaldehyde. After fixation, the intestinal tissue was washed with phosphate-buffered saline solution and then mounted on a slide glass for imaging. All of the experimental procedures and protocols above were conducted in accordance with the guidelines established by the Committee of the Animal Research Policy at Korea University.

**Acknowledgements**

This research was supported by IBS-R023-D1, the National Research Foundation of Korea (NRF) grant funded by the Korea government (MSIT) (No. 2019R1A2C4004804)